# Comparative Analysis of Linepack Impact in Hydrogen and Natural Gas Networks under Dynamic Operating Conditions

Amin Salehi
amin.salehi@aalto.fi

Janne Seppänen
janne.seppanen@aalto.fi

Mahdi Pourakbari-Kasmaei
Mahdi.Pourakbari@aalto.fi

*Department of Electrical Engineering and Automation, Aalto University*, Espoo, Finland.

***Abstract*—Linepack is a critical buffer in gas networks, providing short-term storage and operational flexibility. This paper presents a comparative dynamic analysis of the linepack impact in hydrogen ($H_2$) and natural gas ($CH_4$) networks under compressor contingency conditions. A three-day dynamic simulation is conducted for a $CH_4$ network and a $H_2$ network integrated with a power system. Two cases are investigated: Case 1 with identical and Case 2 with different pipe inner diameters in $CH_4$ and $H_2$ networks. The results in both cases show that $H_2$ exhibits faster transient recovery after the compressor contingency, although its linepack is less than that of $CH_4$. On the other hand, depending on the selected pipe diameter, pressure losses in the two networks can differ significantly. With identical pipe diameters, $H_2$ exhibits lower pressure drop than $CH_4$, and its demand therefore experiences less curtailment. However, when the $H_2$ pipe diameter is reduced, the pressure drop across the $H_2$ pipes increases and the curtailed load becomes higher than in the $CH_4$ network. These results can be used by gas TSOs in designing and operating their grid more efficiently.**

## I. INTRODUCTION

The global transition toward sustainable energy systems has positioned hydrogen ($H_2$) as a promising energy carrier for decarbonizing industrial, commercial, and residential sectors [1], [2]. Since $H_2$ grid relies on intermittent renewable energy sources, flexible energy transport and storage play a vital role in enabling the system to effectively respond to short-term fluctuations in supply and demand. In this context, linepack is one of the most critical operational features of gas networks. Linepack is defined as the inherent storage capacity provided by the pressurized gas within the pipelines. Linepack acts as a crucial buffer to balance short-term imbalances between supply and demand without requiring dedicated external storage facilities [3], [4], [5].

Several recent studies offer $H_2$ blending in the existing natural gas ($CH_4$) pipeline networks as a transitional pathway for decarbonization and hydrogen market development [6], [7], [8]. However, the distinct physical and thermodynamic properties of $H_2$ present unique technical challenges for blending it into the $CH_4$ pipelines [9], [10], and alter the hydraulic response of the network. These changes influence pressure drops, flow dynamics, the amount of useful energy that can be stored within the pipelines, and the rate at which linepack can be increased or decreased under transient conditions. In particular, the lower volumetric energy density of $H_2$ implies that a network converted from $CH_4$ to $H_2$ may exhibit substantially different energy buffering characteristics, even when the physical infrastructure remains unchanged [3], [11]. These issues become especially important under dynamic operation, where injections, withdrawals, and compressor actions vary over time and make pipeline management complex [5]. Therefore, understanding the transient behavior of $H_2$ and $CH_4$ networks is essential for assessing their flexibility in terms of linepack, providing know-hows for future infrastructure planning and operation.

Motivated by the challenges described above, this work compares the linepack impact in $H_2$ and $CH_4$ networks under dynamic conditions. The study investigates how pressure, flow, and stored gas evolve in each network under the contingency of critical components such as compressor stations. Then, it evaluates the extent to which linepack can support operational flexibility in the presence of time-varying fluctuations in supply-demand. The findings aim to clarify the operational differences between $H_2$ and $CH_4$ systems for grid operators and policymakers. In doing so, the paper contributes to a better understanding of how $H_2$ networks should be modeled and operated relative to conventional $CH_4$ grids.

## II. NETWORK MODELING

### A. Natural Gas Network

A $CH_4$ network is designed in SAInt as shown in Fig. 1. SAInt is an integrated platform for planning, modeling, and simulating coupled energy systems such as electricity, gas, and thermal networks [12]. This gas system includes ten nodes, eight pipelines, one $CH_4$ plant, seven demands, and one gas-driven compressor station (GDCS).

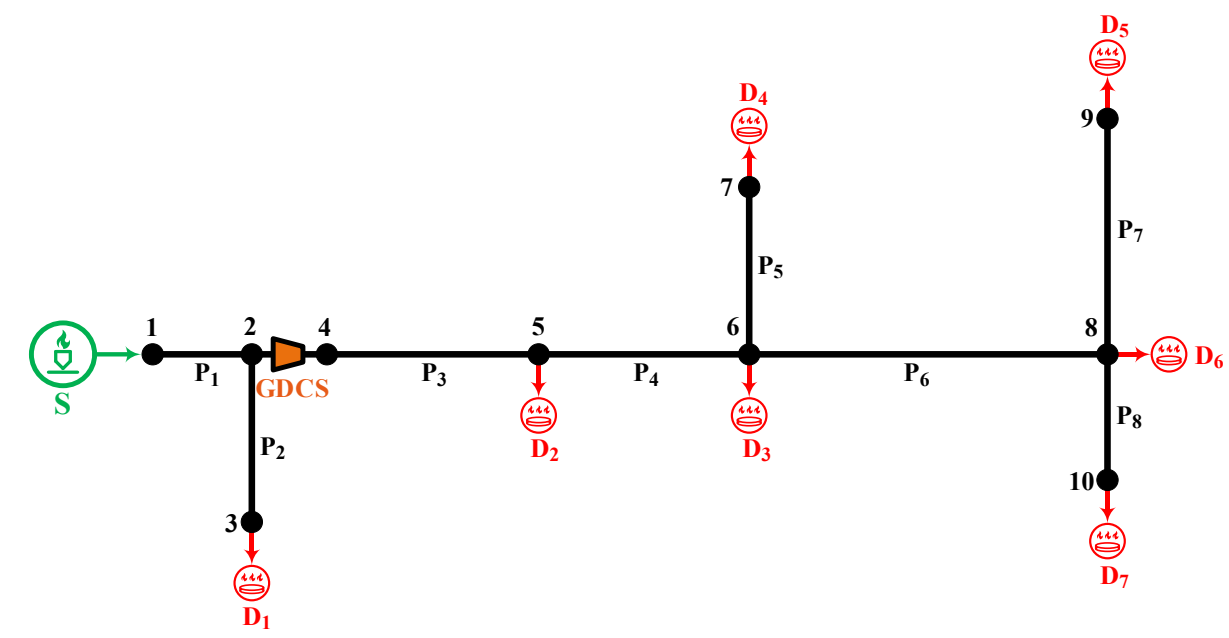

**Fig. 1.** Designed natural gas network.

### B. Hydrogen Network

For the $H_2$ network, the same gas network structure is considered. The $CH_4$ plant and GDCS are replaced with an electrolyzer and an electric-driven compressor station (EDCS), respectively. Additionally, an electricity network is integrated into the corresponding gas network for $H_2$ production. The electric power system includes five nodes, five transmission lines, two generic generators, one wind power plant, and three demands. Two of these electric demands will serve as the power consumption of EDCS's driver and electrolyzer. The integrated network is illustrated in Fig. 2. This hub network is coupled with two connection points, as depicted with dashed blue lines.

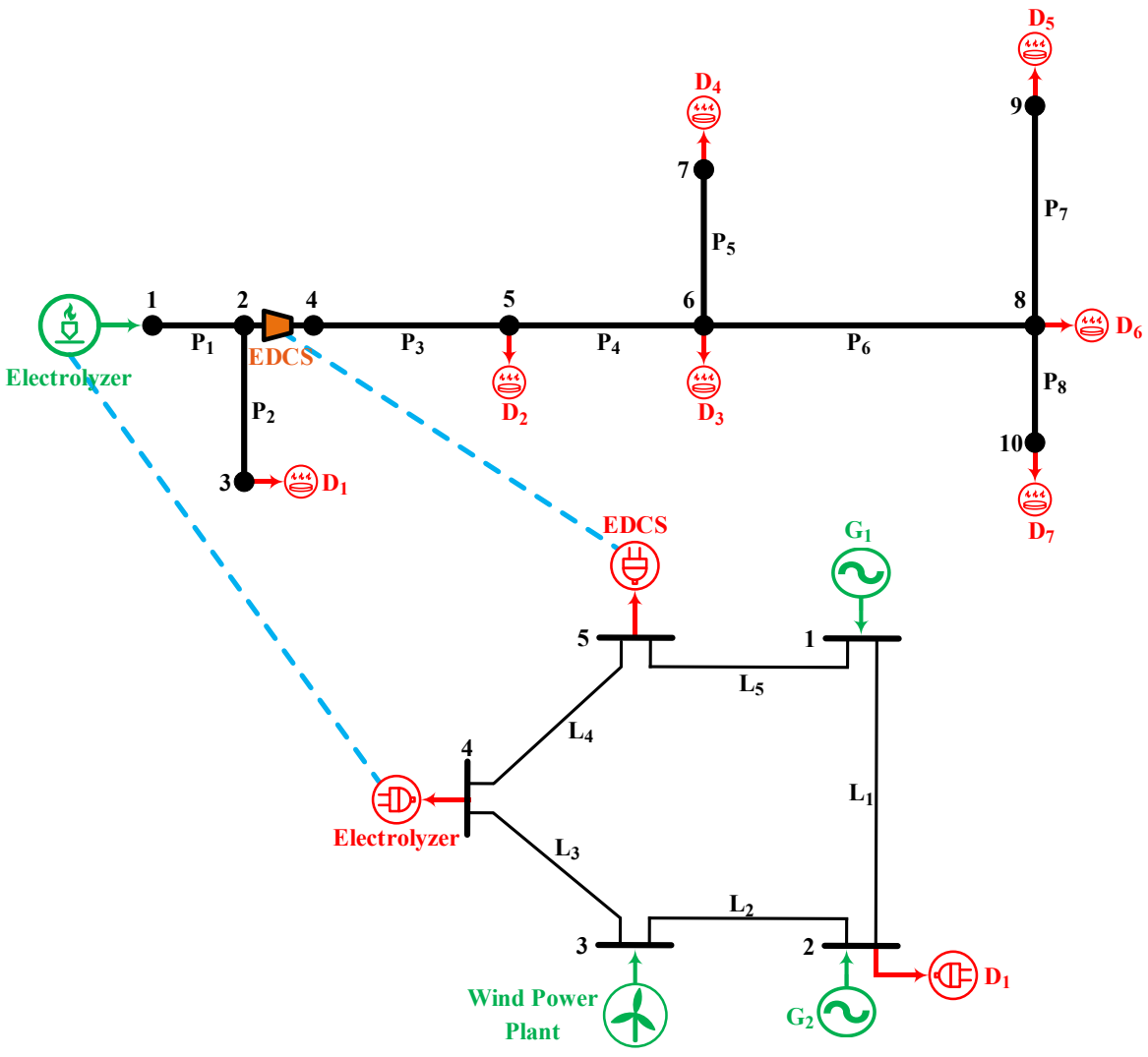

**Fig. 2.** Designed hydrogen network.

## III. Network Data and Case Studies

### A. Natural Gas and Hydrogen Networks Data

Data associated with the $CH_4$ and $H_2$ networks are presented in Tables I-III. In Table I, X and Y represent the coordination of gas nodes. Also, a minimum and maximum pressure is assumed for every node. Accordingly, a penalty price (PP) for all nodes is considered. If maximum pressure associated with a node is violated, the corresponding pressure is set to its maximum, and a penalty will be calculated as ($P$ - $P_{max}$)*PP. Similarly, if minimum pressure associated with a node is violated, the corresponding pressure is set to its minimum, and a penalty will be calculated as ($P_{min}$ - $P$)*PP.

TABLE I. Node data in the $CH_4$ and $H_2$ networks.

| Node | X [km] | Y [km] | $P_{min}$ [bar-g] | $P^{PP}_{min}$ [$/bar] | $P_{max}$ [bar-g] | $P^{PP}_{max}$ [$/bar] |
|---|---|---|---|---|---|---|
| 1 | 0 | 0 | 25 | 100 | 80 | 100 |
| 2 | 25 | 0 | 25 | | 80 | |
| 3 | 25 | -50 | 25 | | 80 | |
| 4 | 25.1 | 0 | 25 | | 80 | |
| 5 | 125.1 | 0 | 25 | | 80 | |
| 6 | 225.1 | 0 | 25 | | 80 | |
| 7 | 225.1 | 50 | 25 | | 80 | |
| 8 | 375.1 | 0 | 40 | | 100 | |
| 9 | 375.1 | 60 | 40 | | 100 | |
| 10 | 375.1 | -40 | 40 | | 100 | |

Table II presents all data associated with the transmission pipes with their specifications. Additionally, the maximum output flow of both gas plants is assumed to be 600 $ksm^3/hr$ and the PP for violating this upper bound is considered 100 $\$/sm^3$. Moreover, Table III summarizes all specifications related to the compressor stations. It is worth noting that the mechanical efficiency for GDCS is considered to be 0.8, while for the EDCS is assumed to be 0.97. This is due to the higher mechanical efficiency of EDCSs compared to GDCS [13].

TABLE II. Pipeline data in the $CH_4$ and $H_2$ networks.

| Pipe | From Node | To Node | L [km] | D* [mm] | WTH* [mm] | RO* [mm] | HTC* [$W/m^2K$] | Eff* |
|---|---|---|---|---|---|---|---|---|
| 1 | 1 | 2 | 25 | 800 | 2 | 0.012 | 0.24 | 1 |
| 2 | 2 | 3 | 50 | 800 | 2 | 0.012 | 0.24 | 1 |
| 3 | 4 | 5 | 100 | 800 | 2 | 0.012 | 0.24 | 1 |
| 4 | 5 | 6 | 100 | 800 | 2 | 0.012 | 0.24 | 1 |
| 5 | 6 | 7 | 50 | 800 | 2 | 0.012 | 0.24 | 1 |
| 6 | 6 | 8 | 150 | 800 | 2 | 0.012 | 0.24 | 1 |
| 7 | 8 | 9 | 60 | 800 | 2 | 0.012 | 0.24 | 1 |
| 8 | 8 | 10 | 40 | 800 | 2 | 0.012 | 0.24 | 1 |

*D: inner diameter; *WTH: wall thickness; *RO: inner pipe roughness; *HTC: heat transfer coefficient; *Eff: pipe efficiency.

TABLE III. EDCS and GDCS data.

| From Node | To Node | $Q_{max}$* [$ksm^3/h$] | $Q^{PP}_{max}$ [$\$/m^3$] | $V_{max}$* [m/s] | $V^{PP}_{max}$ [$/m] | $PD_{max}$* [MW] | $PD^{PP}_{max}$ [$/MWh] |
|---|---|---|---|---|---|---|---|
| 2 | 4 | 500 | 100 | 60 | 100 | 200 | 100 |
| **$PR_{max}$*** | **$PR^{PP}_{max}$ [$]** | **$PI_{min}$* [bar-g]** | **$PI^{PP}_{min}$ [$/bar]** | **$PO_{max}$* [bar-g]** | **$PO^{PP}_{max}$ [$/bar]** | **$\eta_m$*** | **D* [mm]** |
| 5 | 100 | 0 | 100 | 200 | 100 | 0.8 | 800 |

*$Q_{max}$: maximum flow rate; *$V_{max}$: maximum flow velocity; *$PD_{max}$: maximum driver power; *$PR_{max}$: maximum pressure ratio; *$PI_{min}$/$PO_{max}$: minimum/maximum inlet/outlet pressure; *D: inner diameter; *$\eta_m$: driver mechanical efficiency.

### B. Electric Network Data

The data associated with the electrical network are presented in Table IV. This table provides the coordinates of the network nodes as well as the lengths of the transmission lines. The network operates at a base voltage of 400 kV, while the minimum and maximum permissible voltage limits at all nodes are set to 0.95 p.u. and 1.05 p.u., respectively. Additionally, the series impedance and shunt admittance of the transmission lines are assumed to be (0.042 + j0.275) Ω/km and (0 + j4.41) µS/km, respectively. The maximum current capacity and apparent power of all transmission lines are considered to be 1150 A and 1000 MVA, respectively.

TABLE IV. Electric network data.

| Nodes | | | Transmission Lines | | | |
|---|---|---|---|---|---|---|
| **Node** | **X [km]** | **Y [km]** | **Line** | **From Node** | **To Node** | **L [km]** |
| 1 | 300 | -100 | 1 | 1 | 2 | 100 |
| 2 | 300 | -200 | 2 | 2 | 3 | 200 |
| 3 | 100 | -200 | 3 | 3 | 4 | 40 |
| 4 | 70 | -150 | 4 | 4 | 5 | 40 |
| 5 | 100 | -100 | 5 | 5 | 1 | 200 |

### C. Case Studies

A three-day dynamic simulation is conducted in SAInt to analyze the linepack in $H_2$ and $CH_4$ networks. Linepack in steady-state operation conditions can be calculated as (1) [14].

$$LP_{ss} = \frac{\pi D^2 L}{4} \cdot \frac{|P_{avg}| T_b}{P_b T_{avg}} \cdot \frac{1}{Z_{avg}} \quad (1)$$

where, $D$ denotes pipe inner diameter and $L$ presents length. $P_b$, $T_b$, $P_{avg}$, and $T_{avg}$ are the pressure and temperature at standard conditions, and the average pipe pressure and temperature, respectively. $Z_{avg}$ is the average compressibility factor. $P_b$ is supposed to be 1.035 bar-g for both gas systems, and $T_b$ and $T_{avg}$ are assumed to be15° C. The value of $Z_{avg}$ is considered to be 1 for $CH_4$ and 1.06 for $H_2$.

A compressor-station contingency is assumed to occur on the second day of the simulation, starting at 06:00 and lasting until 18:00, for a total duration of 12 hours. Then, two case studies are considered. In Case 1, the $H_2$ and $CH_4$ networks are assumed to have identical pipe inner diameters of 800 mm. In Case 2, different pipe diameters are assigned to the two networks, namely 800 mm for the $CH_4$ network and 600 mm for the $H_2$ network. In both cases, the same volumetric flow rate is assumed to be transported through the pipelines. Nonetheless, designing a $H_2$ pipeline with the same diameter as a $CH_4$ pipeline for the same volumetric flow rate may result in overdesign. Owing to its lower density and different hydraulic characteristics, $H_2$ can be transported through a smaller-diameter pipeline with more flow velocity. In addition, $H_2$ has a lower energy density than $CH_4$, approximately one-third per standard cubic meter [15]. Therefore, for transferring the same volumetric flow, a smaller pipe diameter for $H_2$ transport might be more practical, reducing material requirements and capital costs.

The demands at the load points in the $CH_4$ and $H_2$ networks under all simulation conditions are presented in Table V. It is assumed that all demands have a fixed load profile during simulation period. Accordingly, in the $CH_4$ network, the total demand is defined as the aggregate consumption of all loads and GDCS, which is supplied by the $CH_4$ plant. The plant is assumed to deliver methane at a pressure of 30 bar-g. Then, the GDCS compresses the gas to 55 bar-g before transmitting it through the pipelines. The gas consumption of the GDCS is discussed in Section IV.

In the $H_2$ network, the total $H_2$ production of the electrolyzer is defined as the total consumption of all $H_2$ loads. The power consumption of the EDCS is also assumed to be supplied by the power network. In addition, three loads are considered in the power network: one associated with the electrolyzer plant, one corresponding to the EDCS, and one representing an external power demand. The power consumptions of the electrolyzer and the EDCS are discussed in Section IV, whereas the external power demand is assumed to be a fixed 250 MW load with a power factor of 0.85. Accordingly, the two generic generators and the wind power plant in the power network are assumed to supply these demands. Also, similar to the $CH_4$ network, the electrolyzer plant is assumed to produce $H_2$ at 30 bar-g, after which the EDCS compresses it to 55 bar-g for pipeline transmission.

TABLE V. NATURAL GAS AND HYDROGEN DEMANDS IN [$ksm^3/h$].

| $D_1$ | $D_2$ | $D_3$ | $D_4$ | $D_5$ | $D_6$ | $D_7$ |
|---|---|---|---|---|---|---|
| 80 | 80 | 30 | 50 | 80 | 50 | 80 |

## IV. SIMULATION OUTCOMES

The $CH_4$ network and integrated power and $H_2$ networks are implemented in SAInt simulink environment. The results corresponding to Cases 1 and 2 will be discussed as follows.

*A. Case 1:* Identical pipe diameters in $CH_4$ and $H_2$ networks.

Figure 3 illustrates the operation of the GDCS during the three-day dynamic simulation. The steady-state fuel consumption (FC) of the GDCS is 0.958 $ksm^3/hr$. At this operating point, the compressor is able to compress 370 $ksm^3/hr$ methane from 27.8 bar-g to 55 bar-g, supplying downstream loads. During the contingency, the FC of the compressor drops to zero, leading to a significant pressure decrease on the discharge side and a corresponding pressure increase on the suction side. After the contingency, two transient behaviors are observed. First, the network attempts to return to its steady-state condition immediately after the compressor is restored. This results in a sharp transient peak in FC and discharge pressure, along with a transient drop in suction pressure (see Fig. 3, solid-line boxed region). However, this transient response is constrained by the operational limits of the GDCS. In particular, the maximum allowable flow rate through the compressor is limited to 500 $ksm^3/hr$ (see Table III), which restricts the magnitude of the transient. Subsequently, once the flow rate through the GDCS falls within the allowable limit, the discharge pressure is regulated at 55 bar-g, and the FC gradually returns to its steady-state value (see Fig. 3, dashed-line boxed region). Overall, these dynamic behaviors are consistent with the findings in [16], [17], [18].

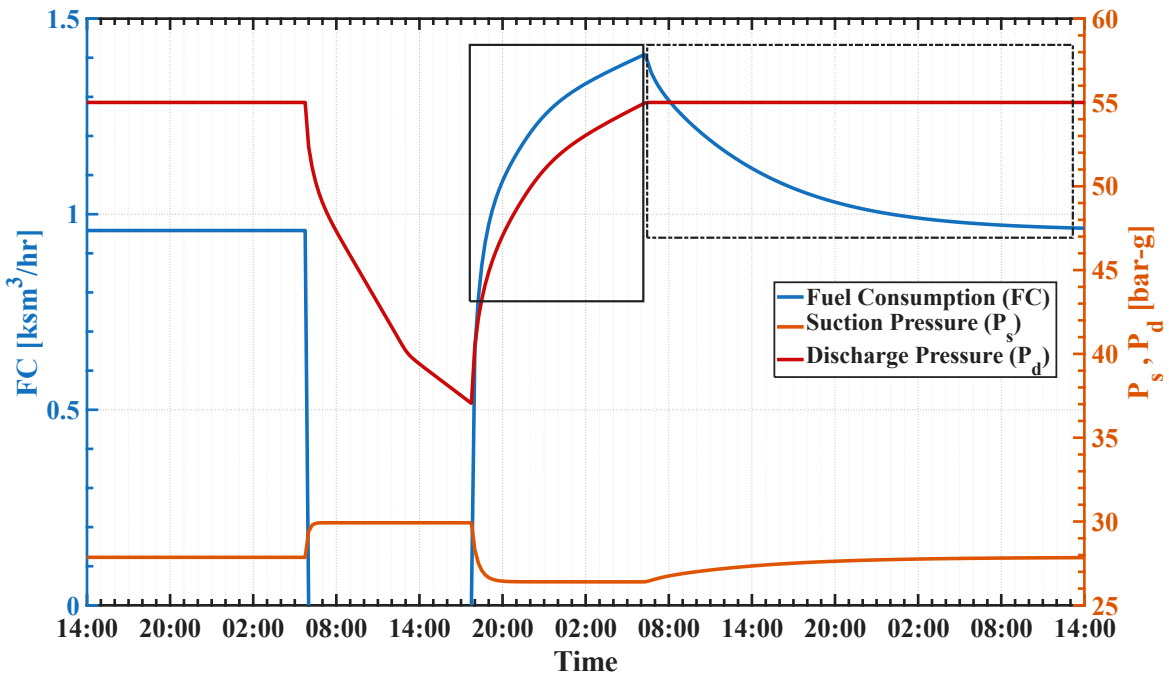

**Fig. 3.** The operation of the GDCS – Case 1.

On the other hand, Fig. 4 demonstrates the operation of the EDCS during the dynamic simulation. The required drive power ($P_D$) of the EDCS is around 8.6 MW to compress 370 $ksm^3/hr$ hydrogen flow from 29.6 bar-g to 55 bar-g. Similar to the GDCS, the same transients occur to the EDCS in the $H_2$ network during and after the compressor contingency.

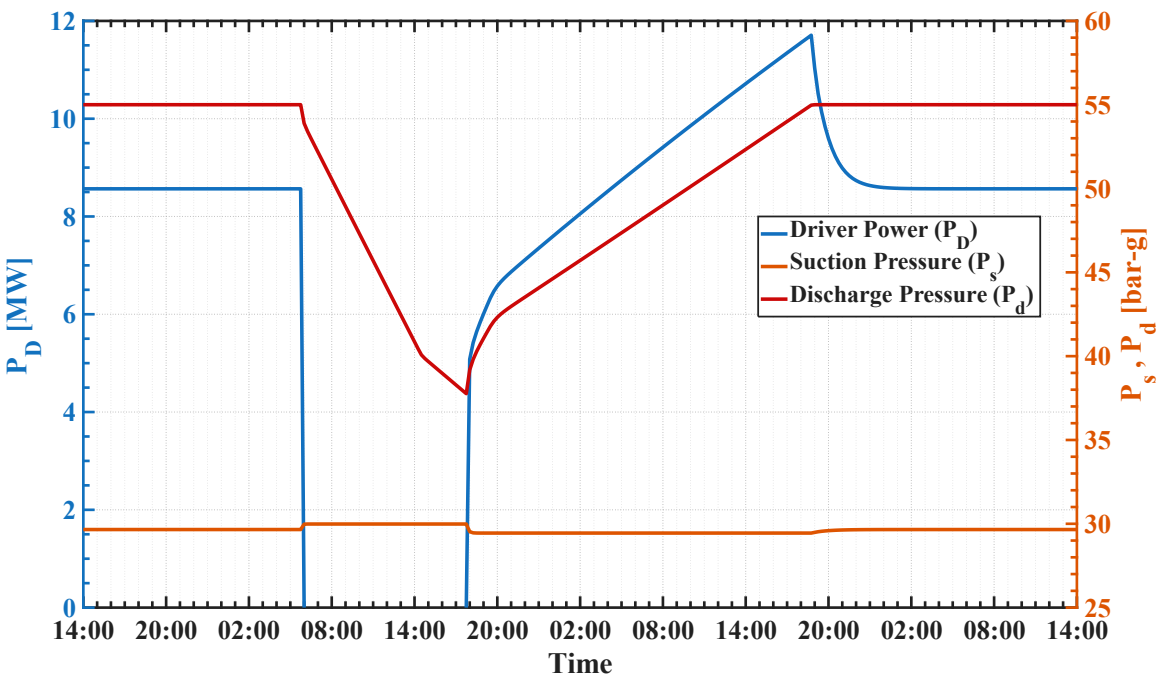

**Fig. 4.** The operation of the EDCS – Case 1.

A comparison of Figs. 3 and 4 indicates that $H_2$ experiences smaller pressure variations than $CH_4$. For instance, the discharge pressure in the EDCS decreases from 55 to 37.75 bar-g during the contingency, whereas in the $CH_4$ case it decreases from 55 to 37 bar-g. Furthermore, the dynamic response of the $H_2$ network is significantly faster than that of the $CH_4$ network. Neglecting the initial transient peaks immediately following the contingency, primarily caused by operational constraints, it is observed that the $H_2$ system variables (e.g., suction and discharge pressures) return to steady-state conditions within approximately 6 hours, while the $CH_4$ system requires more than 24 hours.

Figure 5 demonstrates the production flow rates of $CH_4$ and $H_2$ from the gas plant and electrolyzer, respectively. Under steady-state conditions, both plants operate at a constant flow rate of 450 $ksm^3/hr$, delivering gas at an outlet pressure of 30 bar-g. During the compressor contingency, the production rate of both plants decreases to 80 ksm$^3$/hr, corresponding to the demand of the first load. After the contingency, both the $CH_4$ plant and the electrolyzer exhibit transient behaviors similar to those observed in the GDCS and EDCS. A sharp transient increase in the production flow rate is observed, while this transient response is constrained by the operational limits of the compressors. Hence, the maximum output flow is limited to 580 ksm$^3$/hr, which corresponds to the aggregation of the first demand and the maximum allowable compressor flow rate. Additionally, Fig. 5 clearly indicates that the dynamic response of the $H_2$ system is significantly faster than that of the $CH_4$ system. Transients of $H_2$ annihilate much faster than those in the $CH_4$. It is worth noting that the electrolyzer efficiency is assumed to be 75%. Under this assumption, the electrolyzer requires 2127 MW of active power to produce 450 ksm$^3$/hr of $H_2$. Based on the electrolyzer and EDCS demands, the external power demand, and the transmission-line losses, the active and reactive power outputs of the power plants are presented in Table VI. It is worth noting that the values reported in Table VI correspond to the steady-state operating condition. The power output of all power plants will vary accordingly during the compressor contingency.

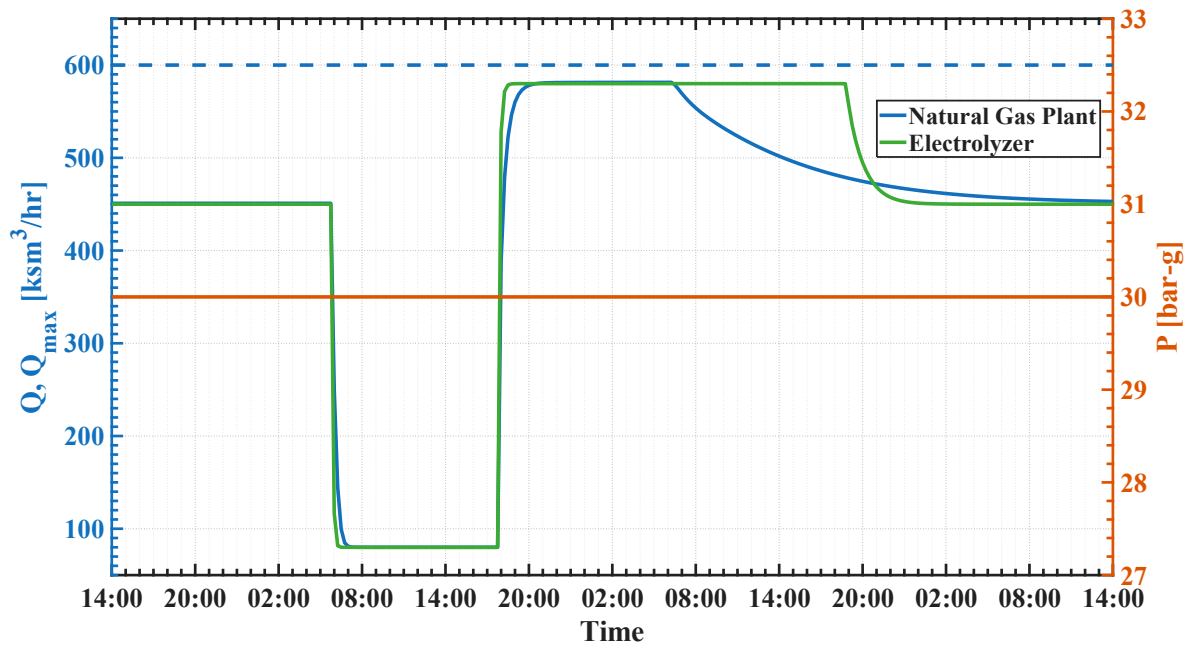

**Fig. 5.** Productions of the natural gas plant and electrolyzer – Case 1.

TABLE VI. ACTIVE AND REACTIVE POWER GENERATION OF POWER PLANTS.

| | $G_1$ | $G_2$ | Wind Power Plant |
|---|---|---|---|
| **P (MW)** | 1018.853 | 804.713 | 700 |
| **Q (MVAr)** | 245.311 | 590.721 | 250 |

Next, the impact of the compressors' contingency on the linepack is investigated. In this regard, the linepack of Pipe 6 (see Fig. 1) is analyzed. This pipeline is critical, as it supplies gas to the loads located at Nodes 8–10. According to Table I, these nodes have higher minimum allowable pressure limits; therefore, their corresponding loads are curtailed if these pressure constraints are not satisfied. Figure 6 presents the linepack of Pipe 6. Under steady-state conditions, the linepack is 3.957 Msm$^3$ and 3.749 Msm$^3$ for $CH_4$ and $H_2$, respectively. Assuming identical standard conditions and pipe inner diameters, the linepack is proportional to the average pipeline pressure and inversely proportional to the compressibility factor, as indicated in (1). The average pressure of Pipe 6 is shown in Fig. 7, where $CH_4$ exhibits a greater pressure drop compared to $H_2$. However, $CH_4$ has a lower compressibility factor than $H_2$. As a result, the overall linepack of $CH_4$ is higher than that of $H_2$. Also, the linepack in both networks decreases significantly during transient conditions. However, the $H_2$ network recovers more rapidly than the $CH_4$ network.

Following the impact of compressors' contingency on the linepack of both networks, it is studied that how this situation affects Demands 5-7. In particular, the pressure at Node 9 is demonstrated in Fig. 8, where it is observed that the pressure in both networks drops below the minimum allowable range. However, the $H_2$ network experiences this nodal pressure drop for a shorter duration. Consequently, the associated load at Node 9 is expected to undergo curtailment for a shorter period. Figure 9 presents the unserved flow of Demand 5 for both networks. In the $CH_4$ network, Demand 5 experiences load shedding for 10.5 hours, with a total curtailed load of 729.3 ksm$^3$. In contrast, the same demand in the $H_2$ network experiences load shedding for 5.25 hours, with a total curtailed load of 409.6 ksm$^3$.

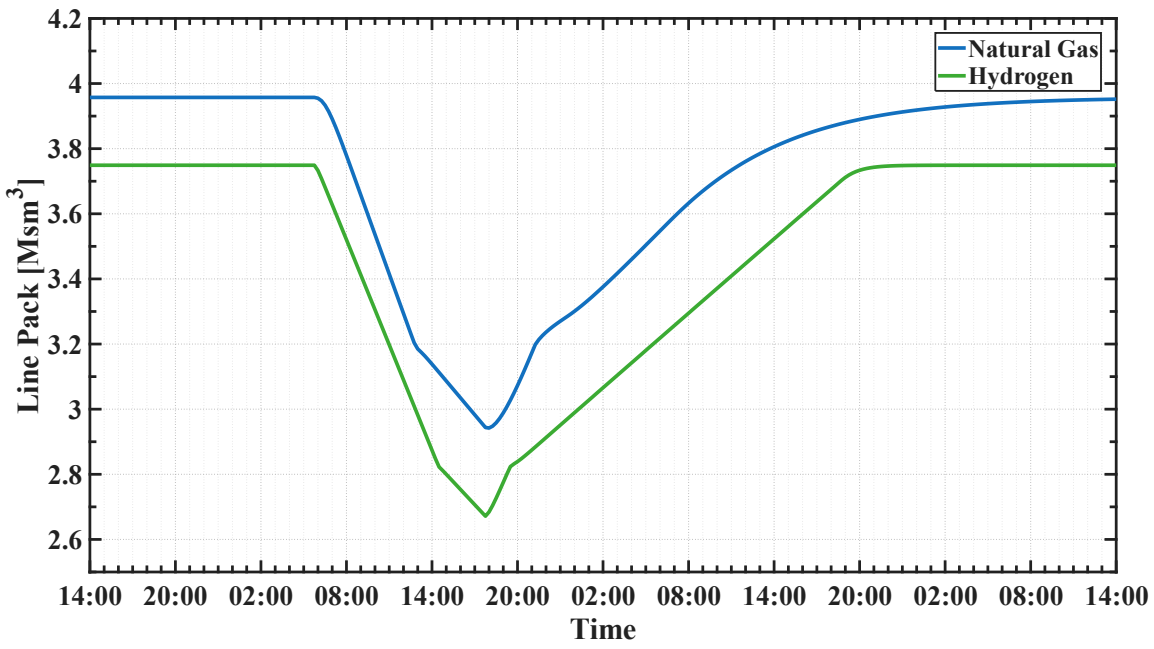

**Fig. 6.** Linepack of Pipe 6 – Case 1.

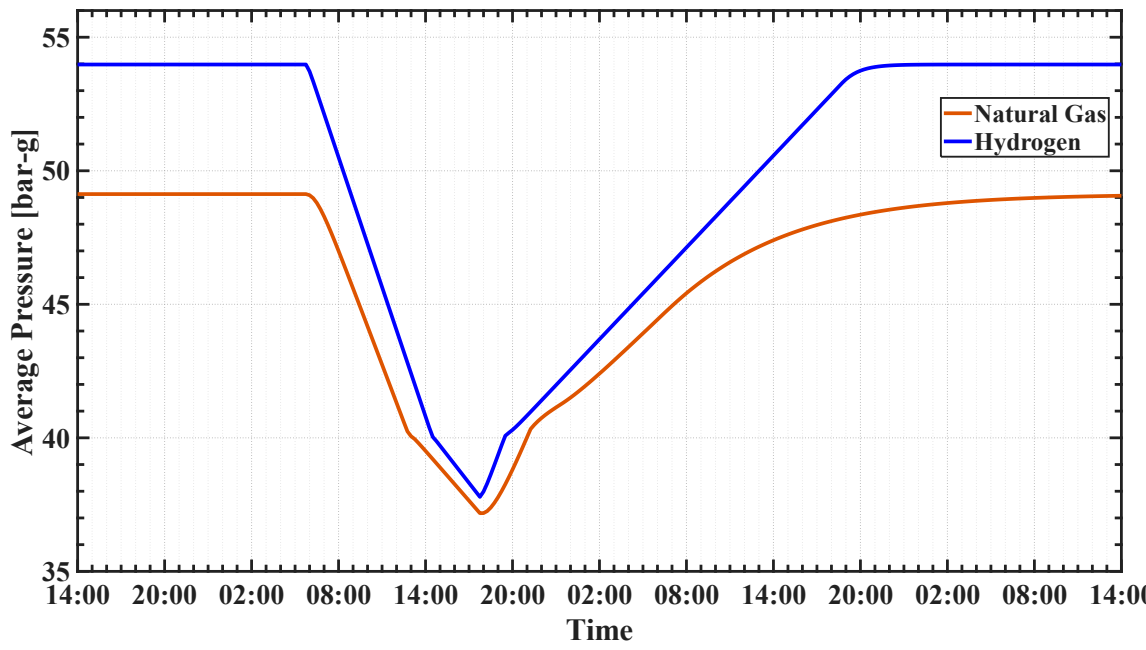

**Fig. 7.** Average pressure of Pipe 6 – Case 1.

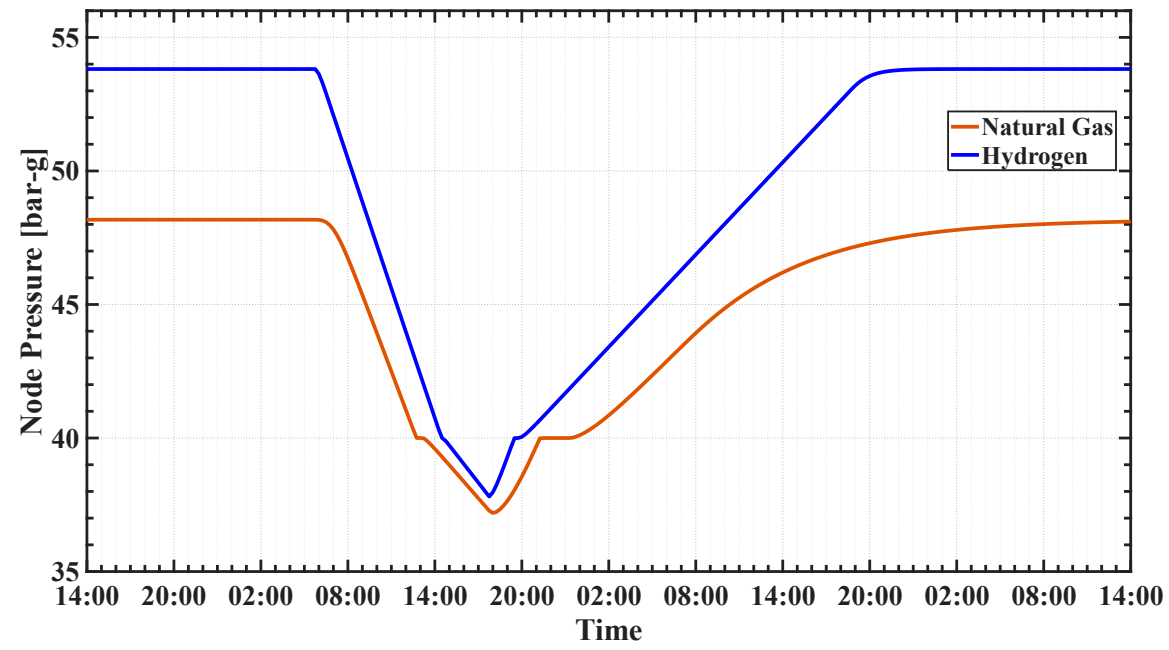

**Fig. 8.** Pressure of Node 9 – Case 1.

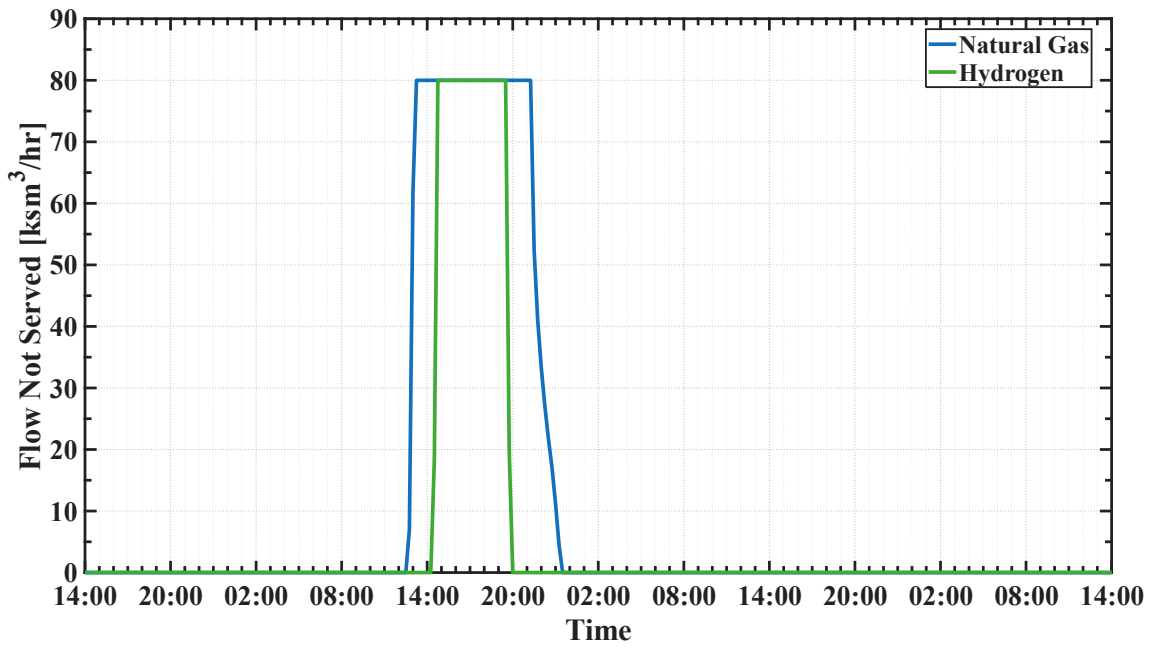

**Fig. 9.** Unserved flow of Demand 5 in Node 9 – Case 1.

The unserved flows of Demands 6 and 7 are also shown in Figs. 10 and 11. All other loads in both networks do not face load shedding as their pressure does not fall below the minimum allowable value during the contingency. Overall, the total curtailed load due to the compressor contingency is 1847.77 ksm$^3$ in the $CH_4$ network and 1060.7 ksm$^3$ in the $H_2$ network. These results indicate that $H_2$ demands experience less curtailment in Case 1. However, it should be noted that the same volumetric flow rate is assumed to be transported in both networks. Therefore, the energy delivered from the two networks should also be compared. For a given volumetric flow rate per hour, the comparison of delivered energy between $CH_4$ and $H_2$ is presented in (2) [19].

$$\frac{E_{H_2}}{E_{CH_4}} = \frac{CV_{H_2}}{CV_{CH_4}} \tag{2}$$

where, CV is the calorific value, which is 39.76 MJ/sm$^3$ for $CH_4$ and 12.76 MJ/sm$^3$ for $H_2$. This comparison shows that methane delivers approximately 3.11 times more energy per unit volume than $H_2$. Therefore, although $H_2$ demands experience less curtailment in Case 1, the amount of energy delivered to these loads is only about one-third of that delivered by $CH_4$ loads.

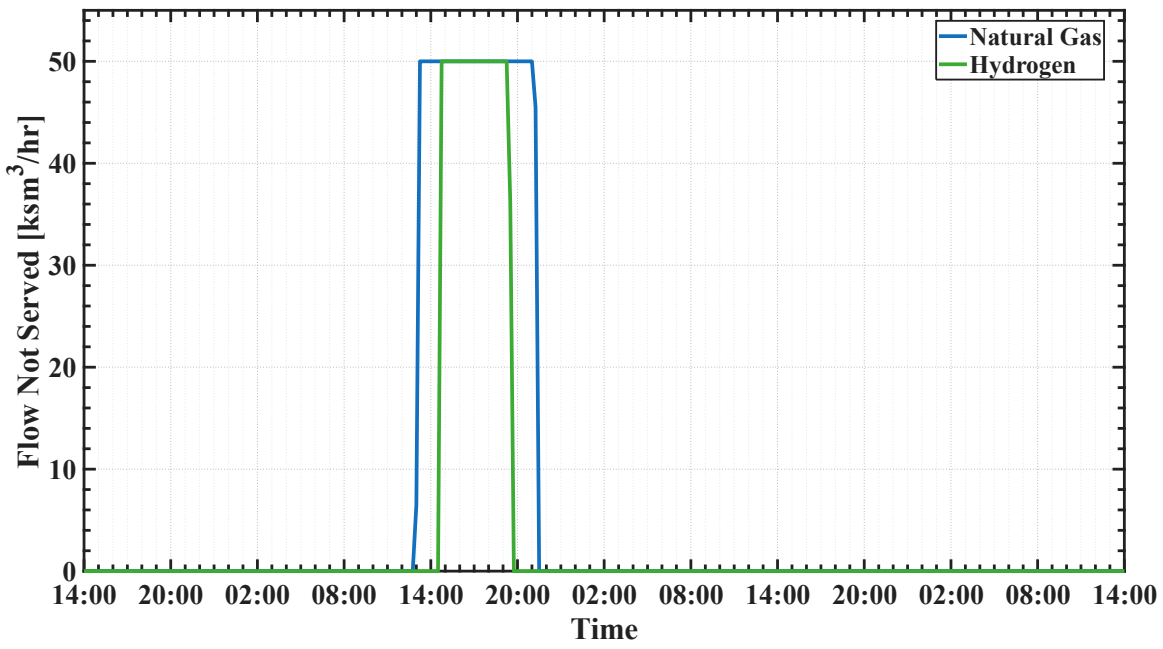

**Fig. 10.** Unserved flow of Demand 6 in Node 8 – Case 1.

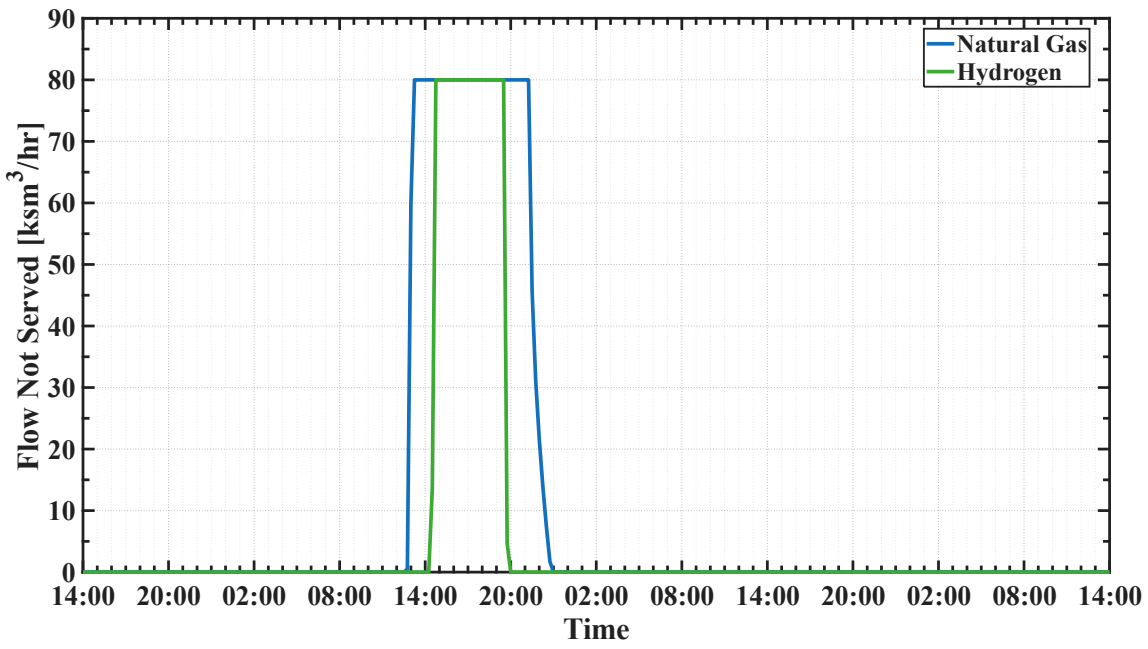

**Fig. 11.** Unserved flow of Demand 7 in Node 10 – Case 1.

*B. Case 2:* Different pipe diameters in $CH_4$ and $H_2$ networks.

In Case 2, similar contingency conditions are examined for the GDCS and EDCS. Nonetheless, a smaller pipe diameter is considered for the $H_2$ network, while the $CH_4$ network retains the same inner diameter as in Case 1. As a result, the operation of the GDCS in this case is identical to that in the previous case and can be traced from Fig. 3. In contrast, the operation of the EDCS differs from Case 1, as illustrated in Fig. 12. In this case, $P_D$ of the EDCS is 9.12 MW to compress 370 ksm$^3$/hr $H_2$ flow from 28.56 bar-g to 55 bar-g. A comparison of Figs. 3 and 12 indicates that $H_2$ experiences larger pressure variations than $CH_4$ in this case. For example, the discharge pressure in the EDCS decreases from 55 to 30.1 bar-g during the contingency, whereas in the GDCS it decreases from 55 to 37 bar-g. Nevertheless, the dynamic response of the $H_2$ network remains faster than that of the $CH_4$ network. Excluding the initial transient peaks immediately after the contingency, the $H_2$ system variables return to steady-state conditions within approximately 12 hours, whereas the $CH_4$ system requires more than 24 hours.

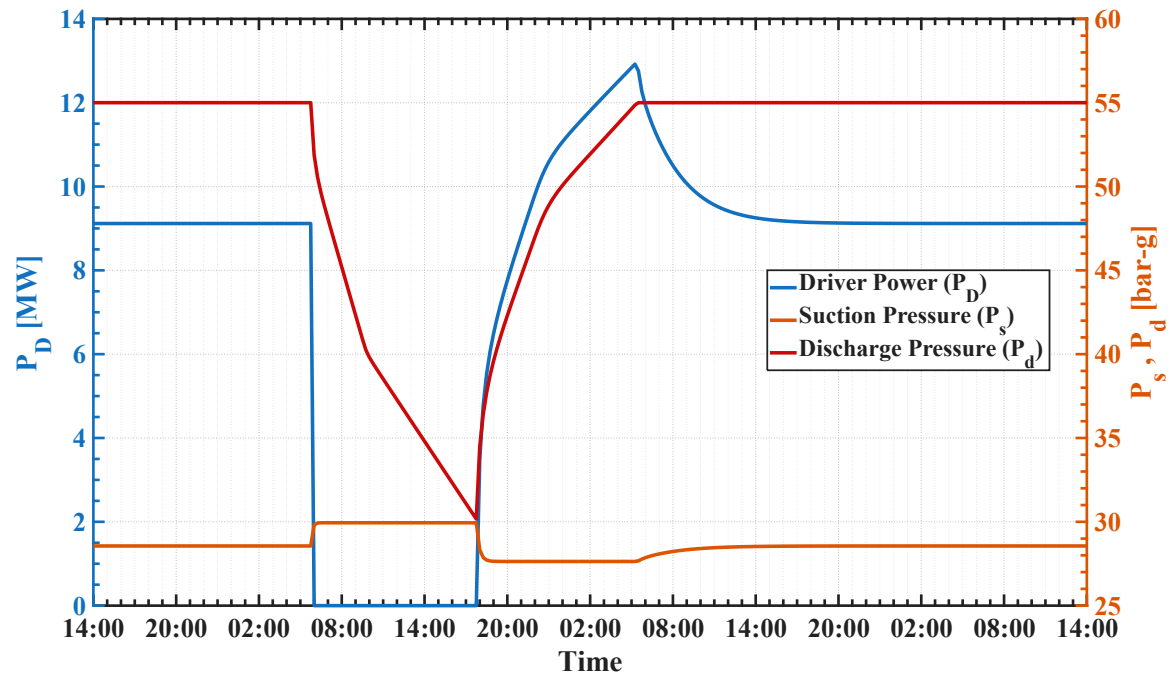

**Fig. 12.** The operation of the EDCS – Case 2.

The production flow rates of the gas plant and electrolyzer are illustrated in Fig. 13. Similar dynamic behaviors to that observed in Case 1 are also present in this case. However, the initial transient response immediately following the contingency is shorter for the electrolyzer in Case 2. Specifically, this transient period lasts approximately 25 hours in Case 1, whereas it is reduced to about 11 hours in Case 2. Nevertheless, the settling time of the $H_2$ network remains shorter than that of the $CH_4$ network. Additionally, the power outputs of the power plants are also the same as those reported in Case 1.

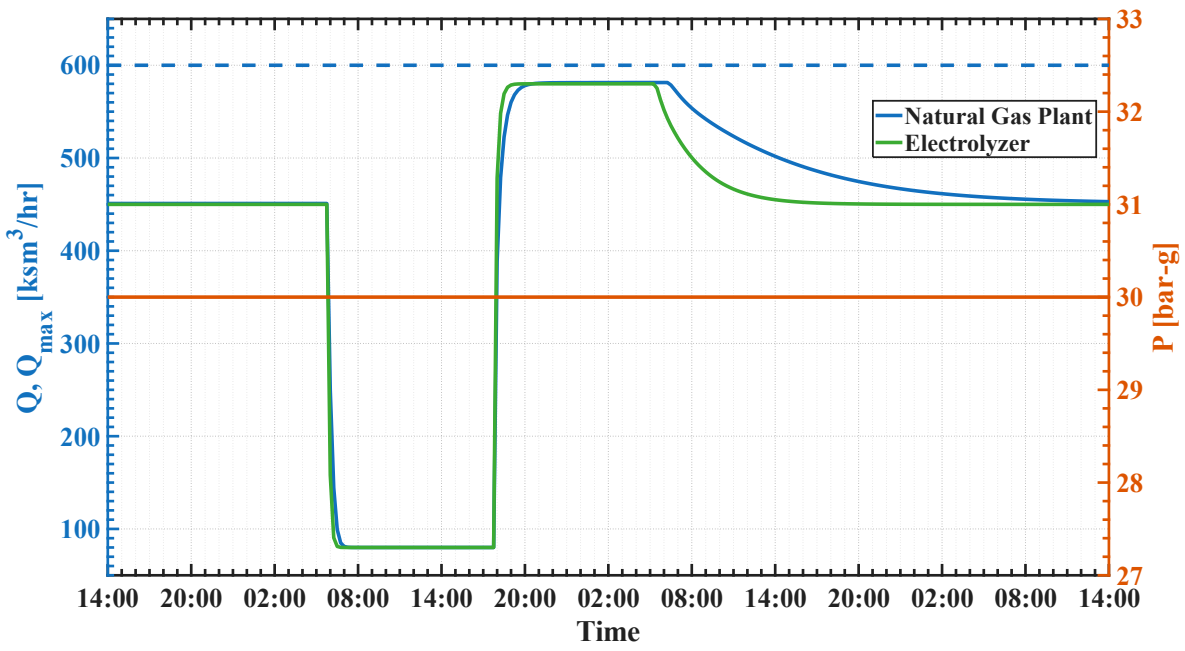

**Fig. 13.** Productions of the natural gas plant and electrolyzer – Case 2.

Similar to the previous case, the impact of the contingency on the linepack of Pipe 6 and the unserved flow of Demands 5-7 is analyzed. Figures 14 and 15 present the linepack of Pipe 6 and its average pressure, respectively. Since the inner diameter of the $H_2$ pipes decreased by 200 mm, the linepack and the average pressure of Pipe 6 decrease drastically. Therefore, the overall linepack of $CH_4$ remains higher than that of $H_2$ in this case. In particular, the linepack of Pipe 6 in the $H_2$ network decreases to 1.987 Msm$^3$, representing a 47 % reduction relative to Case 1.

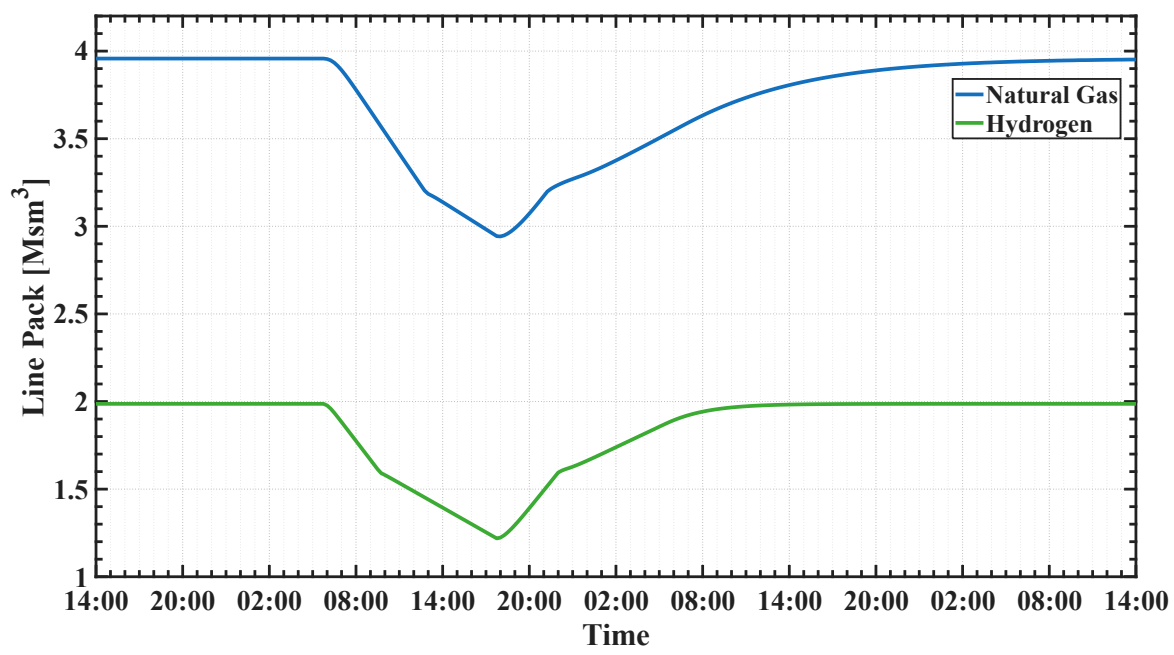

**Fig. 14.** Linepack of Pipe 6 – Case 2.

Additionally, since the pressure loss in the $H_2$ pipes is greater in this case, the loads at Nodes 8-10 are expected to undergo curtailment for a longer period. Figures 16-18 show the unserved flow of Demands 5-7 for both networks, indicating that the $H_2$ demands face more load shedding than the corresponding $CH_4$ demands. Overall, the curtailed loads associated with Demands 5-7 due to the compressor contingency are 612 ksm$^3$, 993.7 ksm$^3$, and 1001.7 ksm$^3$, respectively, giving a total curtailed load of 2607.4 ksm$^3$. In contrast, the total curtailed load in the $CH_4$ network is 1847.77 ksm$^3$.

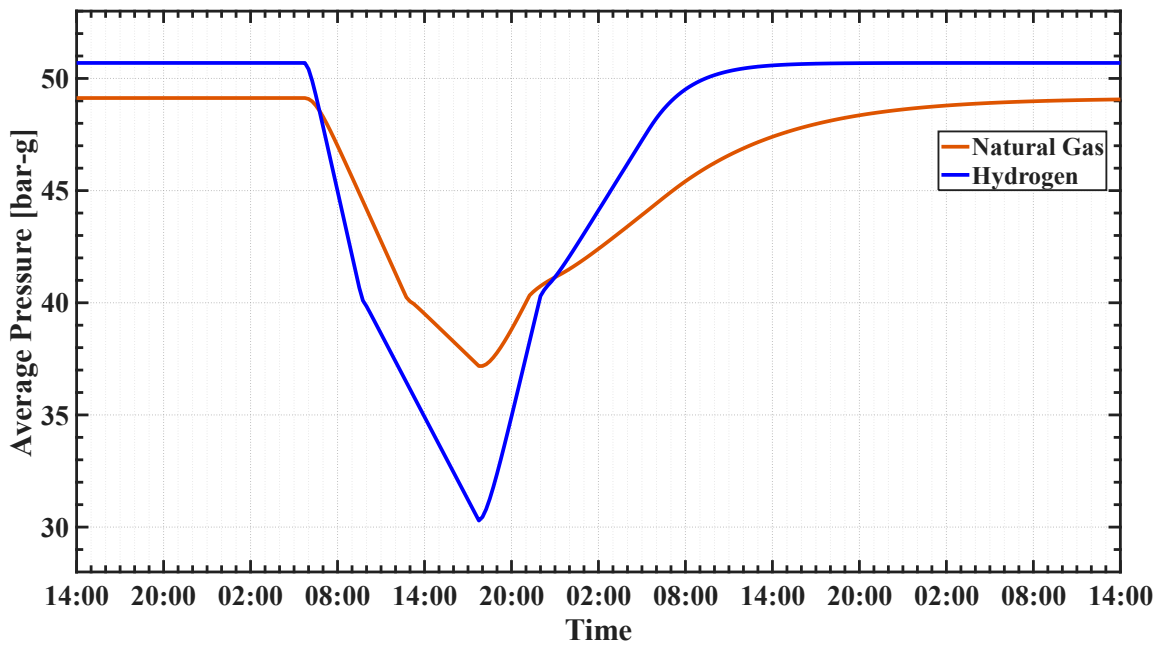

**Fig. 15.** Average pressure of Pipe 6 – Case 2.

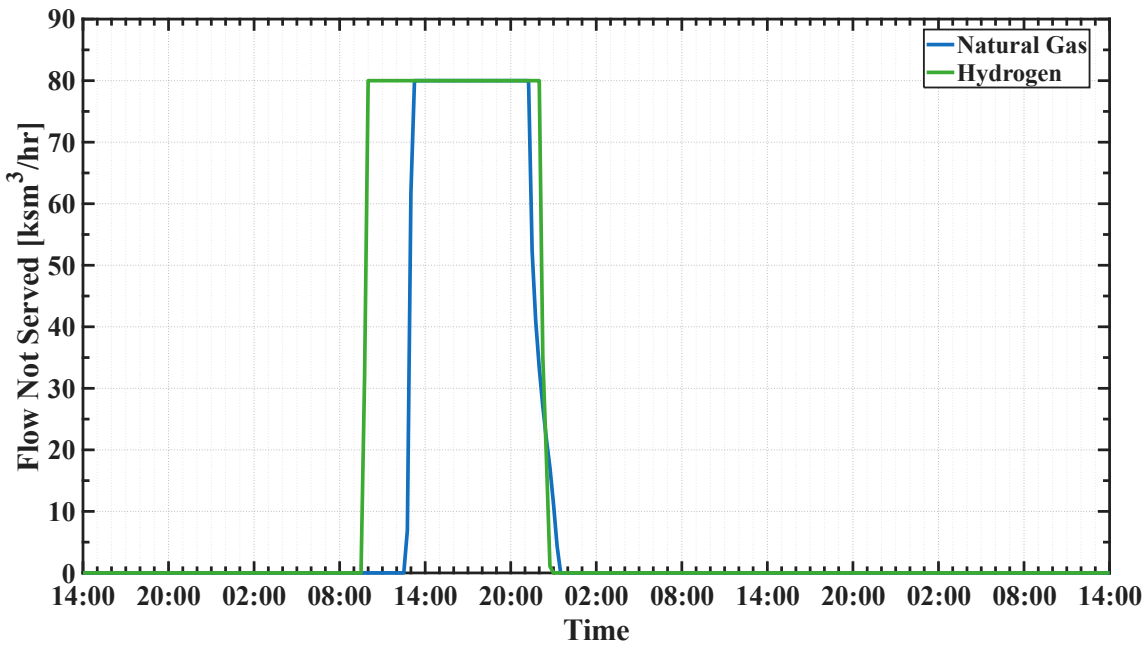

**Fig. 16.** Unserved flow of Demand 5 in Node 9 – Case 2.

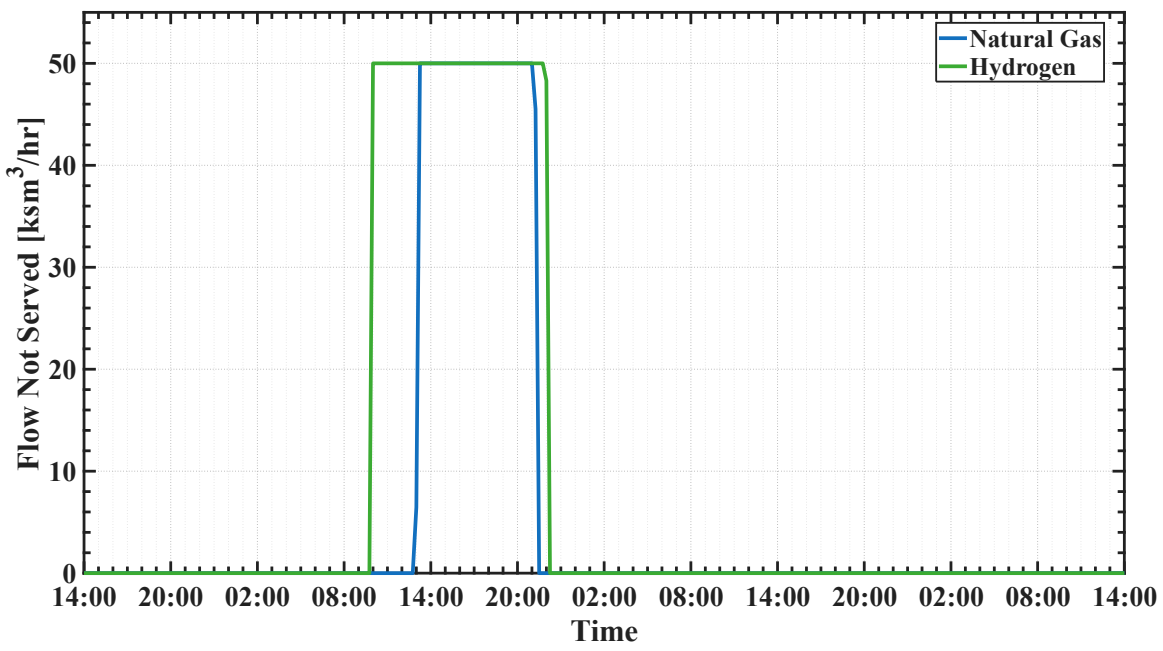

**Fig. 17.** Unserved flow of Demand 6 in Node 8 – Case 2.

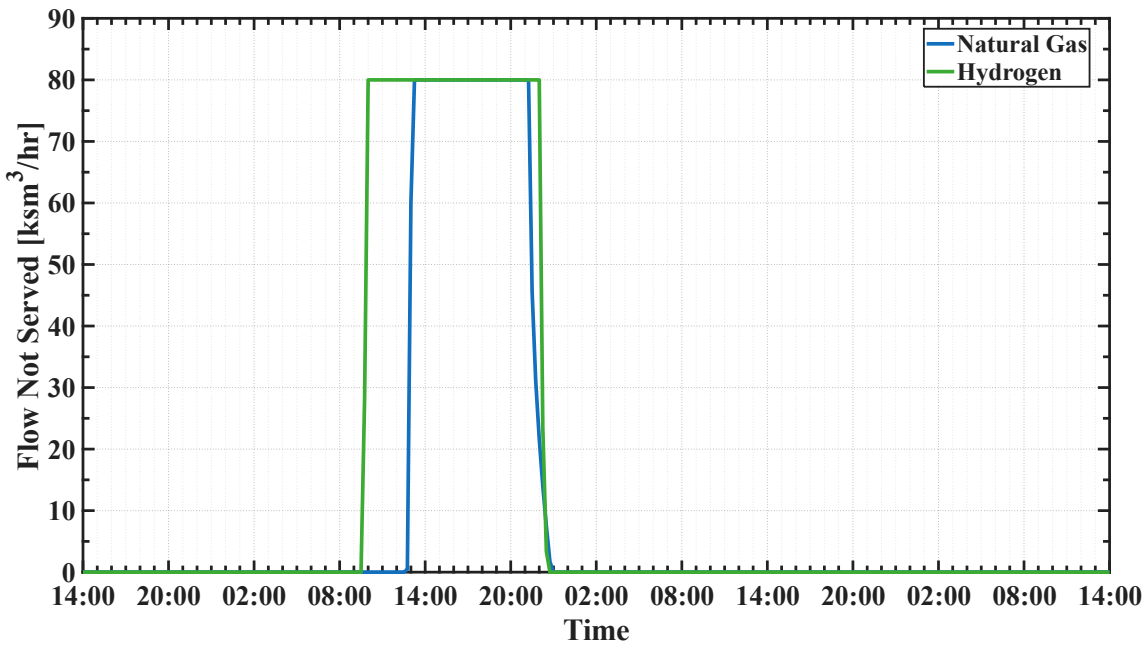

**Fig. 18.** Unserved flow of Demand 7 in Node 10 – Case 2.

## V. CONCLUSION

This study compared the linepack behavior of $H_2$ and $CH_4$ networks under dynamic compressor contingency conditions. The results showed that $H_2$ networks generally recover more rapidly than $CH_4$ networks after a disturbance, but their linepack is less than $CH_4$ for the same volumetric flow rates. In the case of identical pipe diameters, the $H_2$ network exhibited smaller pressure losses and lower load curtailment than the $CH_4$ network. However, because $H_2$ has a much lower volumetric energy content, the amount of energy delivered through the $H_2$ network remained considerably lower for the same transported volumetric flow rate. When a smaller pipe diameter was considered for the $H_2$ network, the pressure drop increased and the linepack capability decreased significantly. This led to greater load curtailment compared with the $CH_4$ network, even though the $H_2$ system still recovered faster from transients. Overall, the findings of this study can support gas TSOs in making more informed decisions regarding the efficient design and operation of gas transmission networks.

## REFERENCES

[1] N. Johnson, M. Liebreich, D. M. Kammen, P. Ekins, R. McKenna, and I. Staffell, “Realistic roles for hydrogen in the future energy transition,” *Nature Reviews Clean Technology*, vol. 1, no. 5, pp. 351–371, 2025.

[2] A. Salehi and M. Pourakbari-Kasmaei, “Wind Farm Capacity Assessment for Hydrogen Production: A Case Study in Finland,” presented at the 2024 IEEE PES Innovative Smart Grid Technologies Europe (ISGT EUROPE), IEEE, 2024, pp. 1–5.

[3] P. Sargent and M. Sargent, “Linepack flexibility for hydrogen and natural gas pipes: A new flexibility metric for energy systems modelling,” *International Journal of Hydrogen Energy*, vol. 132, pp. 139–142, 2025.

[4] T. H. Tran, S. French, R. Ashman, and E. Kent, “Linepack planning models for gas transmission network under uncertainty,” *European Journal of Operational Research*, vol. 268, no. 2, pp. 688–702, 2018.

[5] S. Chevalier and D. Wu, “Dynamic linepack depletion models for natural gas pipeline networks,” *Applied Mathematical Modelling*, vol. 94, pp. 169–186, 2021.

[6] M. A. Kappes and T. E. Perez, “Blending hydrogen in existing natural gas pipelines: Integrity consequences from a fitness for service perspective,” *Journal of Pipeline Science and Engineering*, vol. 3, no. 4, p. 100141, 2023.

[7] M. Davis, A. Okunlola, G. Di Lullo, T. Giwa, and A. Kumar, “Greenhouse gas reduction potential and cost-effectiveness of economy-wide hydrogen-natural gas blending for energy end uses,” *Renewable and Sustainable Energy Reviews*, vol. 171, p. 112962, 2023.

[8] B. C. Erdener *et al.*, “A review of technical and regulatory limits for hydrogen blending in natural gas pipelines,” *International Journal of Hydrogen Energy*, vol. 48, no. 14, pp. 5595–5617, 2023.

[9] N. Rosa, N. A. Fereidani, B. J. Cardoso, N. Martinho, A. Gaspar, and M. G. da Silva, “Advances in hydrogen blending and injection in natural gas networks: A review,” *International Journal of Hydrogen Energy*, vol. 105, pp. 367–381, 2025.

[10] M. Sofian, M. B. Haq, D. Al Shehri, M. M. Rahman, and N. S. Muhammed, “A review on hydrogen blending in gas network: Insight into safety, corrosion, embrittlement, coatings and liners, and bibliometric analysis,” *International Journal of Hydrogen Energy*, vol. 60, pp. 867–889, 2024.

[11] M. Witek and F. Uilhoorn, “Impact of hydrogen blended natural gas on linepack energy for existing high pressure pipelines,” *Archives of Thermodynamics*, pp. 111–124, 2022.

[12] K. A. Pambour, B. C. Erdener, R. Bolado-Lavin, and G. P. Dijkema, “SAInt–A novel quasi-dynamic model for assessing security of supply in coupled gas and electricity transmission networks,” *Applied energy*, vol. 203, pp. 829–857, 2017.

[13] U.S. Environmental Protection Agency (EPA), “Electric Compressor Motors,” Jul. 2024. [Online]. Available: https://tinyurl.com/3shv5tn4

[14] E. S. Menon, *Gas pipeline hydraulics*. Crc Press, 2005.

[15] B. Thawani, R. Hazael, and R. Critchley, “Assessing the pressure losses during hydrogen transport in the current natural gas infrastructure using numerical modelling,” *international journal of hydrogen energy*, vol. 48, no. 88, pp. 34463–34475, 2023.

[16] A. Salehi and M. Pourakbari-Kasmaei, “Dynamics of the Electric-Driven Gas Compressor Station Under Electric and Gas Network Disturbances,” *IEEE Transactions on Industrial Informatics*, 2026.

[17] A. Salehi and M. Pourakbari-Kasmaei, “Electronic Companion: Dynamics of The Electric-Driven Gas Compressor Station Under Electric and Gas Network Disturbances,” Feb. 2026, doi: https://doi.org/10.13140/RG.2.2.35371.30248.

[18] A. Salehi, M. Tofighi-Milani, S. Fattaheian-Dehkordi, M. Lehtonen, J. Seppänen, and M. Pourakbari-Kasmaei, “Investigating the Impact of Electrical Disturbances on the Dynamics of Coupled Electrolyzer-Compressor Systems,” presented at the 2025 IEEE PES Innovative Smart Grid Technologies Conference Europe, IEEE, 2025, pp. 1–5.

[19] O. Massarweh, Y. Bicer, and A. Abushaikha, “Technoeconomic analysis of hydrogen versus natural gas considering safety hazards and energy efficiency indicators,” *Scientific Reports*, vol. 15, no. 1, p. 29601, 2025.